\documentclass[12pt]{article}

\usepackage{amsfonts,amssymb,amsmath}
\usepackage{latexsym}
\usepackage{thm-restate}
\usepackage{url}
\usepackage{cite}

\newcommand{\R}{\mathbb{R}}

\def\01{\{0,1\}}

\newcommand{\norm}[1]{\|#1\|}
\newcommand{\sn}[1]{\|#1\|}
\newcommand{\fro}[1]{\|#1\|_F}
\newcommand{\trn}[1]{\|#1\|_{tr}}

\newcommand{\Tr}{\mathrm{Tr}}
\newcommand{\rk}{\mathrm{rk}}

\newcommand{\braket}[2]{\langle#1, #2\rangle}

\newcommand{\ignore}[1]{}

\newtheorem{theorem}{Theorem}
\newtheorem{lemma}[theorem]{Lemma}
\newtheorem{corollary}[theorem]{Corollary}
\newtheorem{proposition}[theorem]{Proposition}
\newtheorem{definition}[theorem]{Definition}

\newenvironment{proof}[1][Proof: ]
{\noindent {\bf #1}}
{{\hfill $\Box$}\\
 \smallskip}

\begin{document}
\title{An approximation algorithm for approximation rank}
\author{Troy Lee \thanks{Centre for Quantum Software and Information, University of Technology Sydney. Email:troyjlee@gmail.com}
\and
Adi Shraibman \thanks{The Academic College of Tel-Aviv-Yaffo, Tel-Aviv, Israel.  Email: adish@mta.ac.il}
}

\maketitle

\begin{abstract}
One of the strongest techniques available for showing lower bounds on bounded-error communication
complexity is the logarithm of the approximation rank of the communication matrix---the minimum rank
of a matrix which is close to the communication matrix in $\ell_\infty$ norm.  Krause showed
that the logarithm of approximation rank is a lower bound in the randomized case, and later
Buhrman and de Wolf showed it could also be used for quantum communication
complexity.  As a lower bound technique, approximation rank has two
main drawbacks: it is difficult to compute, and it is not known to lower bound the model of quantum
communication complexity with entanglement.

Linial and Shraibman recently introduced a quantity, called $\gamma_2^{\alpha}$,
to quantum communication complexity, showing that it can be used to lower bound communication
in the model with shared entanglement.  Here $\alpha$ is a measure of approximation which is
related to the allowable error probability of the protocol.  This quantity can be written as a
semidefinite program and gives bounds at least as large as many techniques in the literature, although it is smaller than the corresponding $\alpha$-approximation rank, $\rk_\alpha$.  We
show that in fact $\log \gamma_2^{\alpha}(A)$ and $\log \rk_{\alpha}(A)$
agree up to small factors.  As corollaries we obtain a constant factor polynomial time
approximation algorithm to the logarithm of approximation rank, and that the logarithm of
approximation rank is a lower bound for quantum communication complexity with entanglement.
\end{abstract}

\section{Introduction}
Often when trying to show that a problem is computationally hard we ourselves face a
computationally hard problem.  The minimum cost algorithm for a problem is naturally phrased
as an optimization problem, and frequently techniques to lower bound this cost are
also hard combinatorial optimization problems.

When taking such a computational view of lower bounds, it is
natural to borrow ideas from approximation algorithms which have
had a good deal of success in dealing with NP-hardness. Beginning
with the seminal approximation algorithm for MAX CUT of Goemans
and Williamson \cite{GW95}, a now common approach to hard
combinatorial optimization problems is to look at a semidefinite
relaxation of the problem with the hope of showing that such a
relaxation provides a good approximation to the original problem.

We take this approach in dealing with approximation rank, an optimization problem
that arises in communication complexity.  In communication complexity, introduced by Yao \cite{Yao79},
two parties Alice and Bob wish to compute a function $f: X \times Y \rightarrow \{-1,+1\}$, where
Alice receives $x \in X$ and Bob receives $y \in Y$.   The question is how much they have to communicate to evaluate $f(x,y)$ for the most difficult pair $(x,y)$.
Associate to $f$ a $|X|$-by-$|Y|$ {\em communication matrix} $M_f$ where $M_f[x,y]=f(x,y)$.
A well-known lower bound on the deterministic communication complexity of $f$ due to
Mehlhorn and Schmidt \cite{MS82} is $\log \rk(M_f)$.  This lower bound has many nice
features---rank
is easy to compute, at least from a theoretical perspective, and the famous
log rank conjecture of Lov\'{a}sz and Saks \cite{LS88} asserts that this bound is nearly tight
in the sense that there is a universal constant $c$ such that $(\log \rk(M_f))^c$ is an upper
bound on the deterministic communication complexity of $f$, for every function $f$.

When we look at bounded-error randomized communication complexity, where Alice and Bob
are allowed to flip coins and answer incorrectly with some small probability, the relevant quantity is
no longer rank but {\em approximation rank}.  For a sign matrix $A$, the $\alpha$-approximation rank, denoted $\rk_{\alpha}(A)$, is the minimum rank of a matrix $B$ which has the same sign pattern as
$A$ and whose entries have magnitude between $1$ and $\alpha$.  When used to
lower bound randomized communication complexity, the approximation factor
$\alpha$ is related to the allowable error probability of the protocol.  In the limit as
$\alpha \rightarrow \infty$ we obtain the sign rank, denoted $\rk_\infty(A)$, the minimum rank
of a matrix with the same sign pattern as $A$.  Paturi and Simon \cite{PS86} showed that the
$\log \rk_\infty(M_f)$ exactly characterizes the unbounded error complexity of $f$, where
Alice and Bob only have to get the correct answer on every input with probability strictly larger than
$1/2$.  Krause \cite{Kra96} extended this to the bounded-error case by showing that
$\log \rk_{\alpha}(M_f)$ is a lower bound on the $\tfrac{\alpha-1}{2\alpha}$-error randomized
communication complexity of $f$.  Later, Buhrman and de Wolf \cite{BW01} showed that
one-half this quantity is also a lower bound on the bounded-error {\em quantum}
communication complexity of $f$,
when the players do not share entanglement.  Approximation rank is one of the strongest lower
bound techniques available for either of these bounded-error models, giving bounds at
least as large as those given by the discrepancy method, a method
based on Fourier coefficients developed by Raz \cite{Raz95}, and quantum lower methods of Klauck \cite{Kla01} and Razborov \cite{Raz03}.  Notable exceptions include the corruption
bound \cite{Yao83} and information theory methods \cite{CSWY01}, both of which can show an
$\Omega(n)$ lower bound on the communication complexity of disjointness
\cite{Raz92d, BJKS04}, whereas the logarithm of the
approximation rank, and quantum communication complexity, are $\Theta(\sqrt{n})$ for
this problem \cite{Raz03, AA05}.  In view of the log rank conjecture it is natural to conjecture as
well that a polynomial in the logarithm of approximation rank is an upper bound on randomized
communication complexity.

As a lower bound technique, however, approximation rank suffers from two deficiencies.  The
first is that it is quite difficult to compute in practice.  Although we do not know if it is NP-hard
to compute, the class of problems minimizing rank subject to linear constraints does contain
NP-hard instances (see, for example, Section~7.3 in the survey of Vandenberghe and
Boyd \cite{VB96}).  The second drawback is that it is not known to lower bound quantum
communication complexity with entanglement.

We address both of these problems.  We make use of a quantity
$\gamma_2^\alpha$ which was introduced in the context of
communication complexity by Linial et al.\ \cite{LMSS07}.  This
quantity can naturally be viewed as a semidefinite relaxation of
rank, and it is not hard to show that $(\tfrac{1}{\alpha}
\gamma_2^\alpha(A))^2 \le \rk_\alpha(A)$ for a sign matrix $A$
(see Proposition~\ref{prop:rank_lower}).  We show that this lower
bound is in fact fairly tight.
\begin{restatable}{theorem}{mainthm}\label{main}
Let $1 < \alpha < \infty$.  Then for any $m$-by-$n$ sign matrix $A$
\[
\frac{1}{\alpha^2} \ \gamma_2^{\alpha}(A)^2 \le \rk_{\alpha}(A) = 
\frac{2^{16} \alpha^6}{(\alpha-1)^6}
\gamma_2^\alpha(A)^6 \ln^3(8mn) \enspace .
\]
\end{restatable}
The quantity $\gamma_2^\alpha(A)$ can be written as a semidefinite program and so can be
computed up to additive error $\epsilon$ in time polynomial in the size of $A$ and $\log(1/\epsilon)$
by the ellipsoid method (see, for example, the textbook \cite{GLS88}).  Thus Theorem~\ref{main}
gives a constant factor polynomial time approximation algorithm to compute $\log \rk_{\alpha}(A)$.
Moreover, the proof of this theorem gives a method to find a near optimal low rank approximation
to $A$ in randomized polynomial time.

Linial and Shraibman \cite{LS07} have shown that $\log
\gamma_2^{\alpha}(A) - \log \alpha - 2$ is a lower bound on the
$\tfrac{\alpha-1}{2\alpha}$-error quantum communication complexity
of the sign matrix $A$ with entanglement, thus we also obtain the
following corollary.
\begin{corollary}
Let $0 < \epsilon < 1/2$.  Let $Q_{\epsilon}^*(A)$ be the quantum communication complexity of
a $m$-by-$n$ sign matrix $A$ with entanglement.  Then
$$
Q_{\epsilon}^*(A) \ge \frac{1}{6}\log \rk_{\alpha_\epsilon}(A) - \frac{1}{2}\log \log(mn) -
\log\frac{\alpha_\epsilon^2}{\alpha_\epsilon -1} -O(1),
$$
where $\alpha_\epsilon=\tfrac{1}{1-2\epsilon}$.
\end{corollary}
The log log factor is necessary as the $n$-bit equality function with
communication matrix of size $2^n$-by-$2^n$ has
approximation rank $\Omega(n)$ \cite{Alo09}, but can be solved by a bounded-error
quantum protocol with entanglement---or randomized protocol with public coins---with $O(1)$ bits
of communication.  This corollary means that approximation rank cannot be used to show a
large gap between the models of quantum communication complexity with and without entanglement,
if indeed such a gap exists.

Our proof works roughly as follows.  Note that the rank of a $m$-by-$n$ matrix $A$ is the smallest
$k$ such that $A$ can be factored as $A=XY^T$ where $X$ is a $m$-by-$k$ matrix and
$Y$ is a $n$-by-$k$ matrix.  The factorization norm $\gamma_2(A)$ can be defined as
$\min_{X,Y: XY^T=A} r(X) r(Y)$ where $r(X)$ is the largest $\ell_2$ norm of a row of $X$.  Let $X_0,Y_0$ be an optimal solution to this program so
that all rows of $X_0, Y_0$ have
squared $\ell_2$ norm at most $\gamma_2(A)$.
The problem is that, although the rows of $X_0,Y_0$ have small $\ell_2$ norm,
they might still have large {\em dimension}.  Intuitively, however, if the rows of $X_0$ have small
$\ell_2$ norm but $X_0$ has many columns, then one would think that many of the columns are
rather sparse and one could somehow compress the matrix without causing too much damage.  The
Johnson-Lindenstrauss dimension reduction lemma \cite{JL84} can be used to make this
intuition precise.  We randomly project $X_0$ and $Y_0$ to matrices $X_1,Y_1$ with column space of
dimension roughly $\ln(mn) \gamma_2^{\alpha}(A)^2$.  One can argue that with high probability
after such a projection $X_1Y_1^T$ still provides a decent approximation to $A$.  In the second step of
the proof, we do an error reduction step to show that one can then improve this approximation without
increasing the rank of $X_1Y_1^T$ by too much.

Ben-David, Eiron, and Simon \cite{BES02} have previously used this dimension reduction technique to
show that $\rk_\infty(A)=O(\ln(mn) \gamma_2^\infty(A)^2)$ for a sign matrix $A$.  In this limiting case,
however, $\gamma_2^\infty(A)$ fails to be a lower bound on $\rk_\infty(A)$.
Buhrman, Vereshchagin, and de Wolf \cite{BVW07}, and independently Sherstov \cite{She08}, have given an example of a sign matrix $A$ where $\gamma_2^\infty(A)$ is exponentially larger than
$\rk_\infty(A)$.

\section{Preliminaries}
We make use of the Johnson-Lindenstrauss lemma \cite{JL84}.  We state it here in a
form from \cite{BES02} which is most convenient for our use.
\begin{lemma}[Corollary 19, \cite{BES02}]
Let $x,y \in \R^r$. Let $R$ be a random $k$-by-$r$ matrix with entries
independent and identically distributed according to the normal distribution
with mean $0$ and variance $1$.  Then for every $\delta >0$
\[
\Pr_R\left[| \braket{Rx}{Ry} - \braket{x}{y}| \ge \frac{\delta}{2}
\left(\|x\|_2^2+\|y\|_2^2\right)\right] 
\le 4\exp(-\delta^2k/8).
\]
\label{JL_lemma}
\end{lemma}

\subsection{Matrix notation}
We will work with real matrices and vectors throughout this paper.
For a vector $u$, we use $\|u\|$ for the $\ell_2$ norm of $u$, and
$\|u\|_\infty$ for the $\ell_\infty$ norm of $u$. For a matrix $A$,
let $A^T$ denote the transpose of $A$.  We let $A \circ B$ denote
the entrywise product of $A$ and $B$.  We use $S_+^n$ to denote
the set of $n$-by-$n$ symmetric positive semidefinite matrices.
For a symmetric positive semidefinite matrix $M$ let $\lambda_1(M)
\ge \cdots \ge \lambda_n(M) \ge 0$ be the eigenvalues of $M$.  We
define the $i^{th}$ singular value of $A$, denoted $\sigma_i(A)$,
as $\sigma_i(A)=\sqrt{\lambda_i(AA^T)}$.  The rank of $A$, denoted
$\rk(A)$ is the number of nonzero singular values of $A$.  We will
use several matrix norms.
\begin{itemize}
  \item Spectral or operator norm: $\sn{A}=\sigma_1(A)$.
  \item Trace norm: $\trn{A}=\sum_i \sigma_i(A)$.
  \item Frobenius norm: $\fro{A}=\sqrt{\sum_i \sigma_i(A)^2}$.
\end{itemize}
One can alternatively see that $\fro{A}^2=\Tr(AA^T)=\sum_{i,j} A[i,j]^2$.

Our main tool will be the factorization norm $\gamma_2$ \cite{TJ89}, introduced in the context of
complexity measures of matrices by Linial et al.\ \cite{LMSS07}.  This norm can naturally be viewed as a
semidefinite programming relaxation of rank as we now explain.
We take the following as our primary definition of $\gamma_2$:
\begin{definition}[\cite{TJ89,LMSS07}]
Let $A$ be a $m$-by-$n$ matrix.  Then
$$
\gamma_2(A)=\min_{X,Y: X Y^T=A} r(X) r(Y),
$$
where $r(X)$ is the largest $\ell_2$ norm of a row of $X$.
\label{gamma_primal_def}
\end{definition}

We can write $\gamma_2(A)$ as the optimum value of a semidefinite program as follows.
\begin{align*}
\gamma_2(A)&= \min_{P \in S_+^{m+n}} c \\
                         &P[i,i] \le c \mbox{ for all } i=1, \ldots, m+n \\
                         & P[i,j+m]=A[i,j] \mbox{ for } i=1, \ldots, m, j=1, \ldots,n \\
\end{align*}
This is because given a factorization $XY^T=A$, we can create a positive semidefinite matrix
$$
P=
\begin{pmatrix}
XX^T & XY^T \\
YX^T & YY^T
\end{pmatrix}
$$
satisfying the constraints of this semidefinite program.  Conversely, given a positive semidefinite
matrix $P$ satisfying the constraints of the program, we can write $P=ZZ^T$ and let
$X$ be the first $m$ rows of $Z$ and $Y$ the last $n$ rows of $Z$ to obtain a factorization of
$A=XY^T$.

The quantity $\gamma_2$ can equivalently be written as the optimum of a maximization problem
known as the Schur product operator norm: $\gamma_2(A)=\max_{X: \|X\|=1} \|A \circ X\|$.
The book of Bhatia (Thm. 3.4.3 \cite{Bha07}) contains a nice discussion of this equivalence
and attributes it to an unpublished manuscript of Haagerup.  An alternative proof can be
obtained by dualizing the above semidefinite program \cite{LSS08}.

More convenient for our purposes will be a formulation of $\gamma_2$ in terms of the trace norm.
One can see that this next formulation is equivalent to the Schur product operator norm formulation
using the fact that $\trn{A}=\max_{B: \sn{B} \le 1} \Tr(AB^T)$.

\begin{proposition}[cf. \cite{LSS08}]
Let $A$ be a matrix.  Then
$$
\gamma_2(A)=\max_{\substack{u,v \\ \norm{u}=\norm{v}=1}} \trn{A \circ vu^T}
$$
\label{prop:rank_lower}
\end{proposition}

From this formulation we can easily see the connection of $\gamma_2$ to matrix rank.
This connection is well known in Banach spaces theory, where it is proved in a more
general setting, but the following proof is more elementary.

\begin{proposition}[\cite{TJ89,LSS08}]
\label{gamma_2_rank}
Let $A$ be a matrix.  Then
$$
\rk(A) \ge \frac{\gamma_2(A)^2}{\|A\|_\infty^2}.
$$
\end{proposition}

\begin{proof}
Let $u,v$ be unit vectors such that $\gamma_2(A)=\trn{A \circ vu^T}$.  As the rank of $A$ is equal
to the number of nonzero singular values of $A$, we see by the Cauchy-Schwarz inequality that
$$
\rk(A) \ge \frac{\trn{A}^2}{\fro{A}^2}.
$$
As $\rk(A \circ vu^T) \le \rk(A)$ we obtain
\begin{align*}
\rk(A) &\ge \frac{\trn{A \circ vu^T}^2}{\fro{A \circ vu^T}^2} \\
&\ge \frac{\gamma_2(A)^2}{\|A\|_\infty^2}
\end{align*}

\end{proof}

Finally, we define the approximate version of the $\gamma_2$ norm.
\begin{definition}[\cite{LS07}]
Let $A$ be a sign matrix, and let $\alpha \ge 1$.
\begin{align*}
\gamma_2^{\alpha}(A)&=\min_{B: 1 \le A[i,j] B[i,j] \le \alpha} \gamma_2(B) \\
\gamma_2^{\infty}(A)&=\min_{B: 1 \le A[i,j] B[i,j]} \gamma_2(B)
\end{align*}
\end{definition}
We define approximation rank similarly.
\begin{definition}[approximation rank]
Let $A$ be a sign matrix, and $\alpha \ge 1$.
\begin{align*}
\rk_{\alpha}(A)&=\min_{B: 1 \le A[i,j] B[i,j] \le \alpha} \rk(B) \\
\rk_{\infty}(A)&=\min_{B: 1 \le A[i,j] B[i,j]} \rk(B)
\end{align*}
\end{definition}

As corollary of Proposition~\ref{gamma_2_rank} we get
\begin{corollary}
Let $A$ be a sign matrix and $\alpha \ge 1$.
$$
\rk_\alpha(A) \ge \frac{1}{\alpha^2} \ \gamma_2^\alpha(A)^2
$$
\end{corollary}

\section{Main Result}
In this section we present our main result relating $\gamma_2^\alpha(A)$ and $\rk_{\alpha}(A)$.
We show this in two steps: first using dimension reduction we upper bound $\rk_{\alpha'}(A)$ in terms of
$\gamma_2^{\alpha}(A)$ where $\alpha'$ is slightly larger than $\alpha$.  In the second step of error reduction
we show how to decrease the error back to $\alpha$ without increasing the rank too much.

\subsection{Dimension reduction}

\begin{theorem}
\label{d_reduc}
Let $A$ be a $m$-by-$n$ sign matrix and $\alpha \ge 1$.  Then for any $0 < t < 1$
$$
\rk_{\frac{\alpha+t}{1-t}}(A) \le \frac{8\gamma_2^{\alpha}(A)^2 \ln(8mn)}{t^2}
$$
\end{theorem}
\begin{proof}
Suppose that $\gamma_2^\alpha(A)=\gamma$.  By the formulation in
Definition~\ref{gamma_primal_def},
this means there is a set of vectors $x_i \in \R^r$ for $i=1, \ldots, m$ and
$y_j \in \R^r$ for $j=1, \ldots, n$ such that
\begin{itemize}
  \item $1 \le \braket{x_i}{y_j}A[i,j] \le \alpha$ for all $i,j$
  \item $\|x_i\|^2, \|y_j\|^2 \le \gamma$ for all $i,j$.
\end{itemize}
Applying the Johnson-Lindenstrauss lemma (Lemma~\ref{JL_lemma}) with $\delta = t/\gamma$ we have
\begin{align*}
\Pr_R\left[| \braket{Rx_i}{Ry_j} - \braket{x_i}{y_j}| \ge  t \right]
\le 4\exp\left(\frac{-t^2 k}{8\gamma^2}\right) \enspace,
\end{align*}
where the probability is taken over $k$-by-$r$ matrices with each entry chosen independently and identically distributed according to the 
standard normal distribution.  By taking
$k=8\gamma^2 \ln(8mn)/t^2$ we can make the failure probability at most $1/(2mn)$.  Thus by a union bound
we have that $|\braket{Rx_i}{Ry_j}-\braket{x_i}{y_j}| \le t$ for all $i,j$ with probability
at least $1/2$ over the choice of $R$.  Thus there exists a matrix $R_0$ where this holds and by defining the $m$-by-$n$ matrix $B$ where 
$B(i,j) = \frac{1}{1-t} \braket{R_0x_i}{R_0y_j}$, we see that $B$ has rank at most $k$ and gives an $\frac{\alpha+t}{1-t}$-approximation 
to $A$.
\end{proof}

\subsection{Error-reduction}
In this section, we will see how to improve the approximation factor a matrix $A'$ gives to
a sign matrix $A$ without increasing its rank by too much.  We do this by applying a
low-degree polynomial approximation of the sign function to the entries of $A'$.  This
technique has been used several times before.  The trick of controlling
the value of inner products $\braket{x}{y}$ by taking (sums of) tensor products of $x$ and $y$
can be found in Krivine's proof of Grothendieck's inequality \cite{Kri79};
results more specifically related to our context can be found, for example, in \cite{Alo03, KS07}.

We first need a lemma of Alon \cite{Alo03} about how applying a degree $d$ polynomial
to a matrix entrywise can increase its rank.  For completeness we give the proof of
a weaker version of this lemma here.  Let $p(x)=a_0 + a_1 x + \ldots + a_d x^d$ be a degree
$d$ polynomial.  For a matrix $A$, we define $p(A)$ to be the matrix
$a_0 J + a_1 A + \ldots + a_d A^{\circ d}$ where $A^{\circ s}$ is the matrix whose $(i,j)$ entry is
$A[i,j]^s$, and $J$ is the all ones matrix.

\begin{lemma}
\label{rk_error_reduc}
Let $A$ be a matrix and $p$ be a degree $d$ polynomial.  Then
$\rk(p(A)) \le (d+1) \rk(A)^d$
\end{lemma}

\begin{proof}
The result follows using subadditivity of rank and that rank is multiplicative under tensor
product.  We have $\rk(A^{\circ s}) \le \rk(A^{\otimes s})=\rk(A)^s$
since $A^{\circ s}$ is a submatrix of $A^{\otimes s}$.
\end{proof}

In general for any constants $1< \beta \le \alpha < \infty$ one can show that there is a constant
$c$ such that $\rk_\beta(A) \le \rk_\alpha(A)^c$ by looking at low degree approximations of the
sign function (see Corollary~1 of \cite{KS07} for such a statement).  As we are interested in the
special case where $\alpha,\beta$ are quite close, we give an explicit construction in an attempt
to keep the exponent as small as possible.

\begin{proposition}
\label{prop:poly}
Fix $\epsilon >0$.  Let $a_3=1/(2+6\epsilon+4\epsilon^2)$, and $a_1=1+a_3$.  Then the
polynomial
$$
p(x)=a_1 x -a_3x^3
$$
maps $[1,1+2\epsilon]$ into $[1,1+\epsilon]$ and $[-1-2\epsilon,-1]$ into $[-1-\epsilon,-1]$.
\end{proposition}

\begin{proof}
As $p$ is an odd polynomial, we only need to check that it maps $[1, 1+2\epsilon]$ into
$[1, 1+\epsilon]$.  With our choice of $a_1, a_3$, we see that $p(1)=p(1+2\epsilon)=1$.
Furthermore, $p(x) \ge 1$ for all $x \in [1, 1+2\epsilon]$, thus we just need to check that
the maximum value of $p(x)$ in this interval does not exceed $1+\epsilon$.

Calculus shows that the maximum value of $p(x)$ is attained at $x=(\tfrac{1+a_3}{3a_3})^{1/2}$.
Plugging this into the expression for $p(x)$, we see that the maximum value is
$$
\max_{x \in [1, 1+2\epsilon]} p(x) = \frac{2}{3\sqrt{3}} \frac{(1+a_3)^{3/2}}{\sqrt{a_3}}.
$$
We want to show that this is at most $1+\epsilon$, or equivalently that
$$
\frac{2}{3\sqrt{3}} \frac{\sqrt{2+6\epsilon+4\epsilon^2}}{1+\epsilon}\left(\frac{3+6\epsilon+4\epsilon^2}
{2+6\epsilon+4\epsilon^2}\right)^{3/2} \le 1.
$$
One can verify that this inequality is true for all $\epsilon \ge 0$.
\end{proof}

\subsection{Putting everything together}
Now we are ready to put everything together.
\mainthm*

\begin{proof}
We first apply Theorem~\ref{d_reduc} with $t = \frac{\alpha-1}{2\alpha}$.  With this choice, $\frac{\alpha+t}{1-t} = 2\alpha-1$, 
thus we obtain
\[
\rk_{2\alpha-1}(A) \le \frac{32\alpha^2}{(\alpha-1)^2} \gamma_2^{\alpha}(A)^2 \ln(8mn) \enspace .
\]
Now we can use the polynomial constructed in Proposition~\ref{prop:poly} and
Lemma~\ref{rk_error_reduc} to obtain
\[
\rk_\alpha(A) \le 2\rk_{2\alpha-1}(A)^3 \le \frac{2^{16} \alpha^6}{(\alpha-1)^6}
\gamma_2^\alpha(A)^6 \ln^3(8mn)  \enspace.
\]
\end{proof}

\section{Discussion and open problems}
One of the fundamental questions of quantum information is the power of entanglement.  If we
believe
that there can be a large gap between the communication complexity of a function with and without
entanglement then we must develop techniques to lower bound quantum communication
complexity
without entanglement that do not also work for communication complexity with entanglement.  We
have eliminated one of these possibilities in approximation rank.

As can be seen in Theorem~\ref{main}, the relationship between
$\gamma_2^\alpha(A)$ and $\rk_\alpha(A)$ weakens as $\alpha \rightarrow \infty$ because
the {\em lower bound} becomes
worse.  Indeed, Buhrman, Vereshchagin, and de Wolf \cite{BVW07}, and independently Sherstov
\cite{She08}, have given examples where $\gamma_2^\infty(A)$ is exponentially larger than
$\rk_\infty(A)$.  It is an interesting open problem to find a polynomial time approximation algorithm
for the sign rank $\rk_\infty(A)$.  It is known that the sign rank itself is
NP-hard to compute \cite{BFGJK09,BK15}.

\section*{Acknowledgments}
We would like to thank Ronald de Wolf for helpful comments on an earlier 
version of this manuscript and Gideon Schechtman for helpful conversations.
We also thank Shalev Ben-David for pointing out an error in a previous version of 
the proof of Theorem~\ref{d_reduc}.
This work conducted while TL was at Rutgers University, supported by a 
NSF mathematical sciences postdoctoral fellowship.

\newcommand{\etalchar}[1]{$^{#1}$}

\end{document}